\documentclass[a4paper]{article}

\usepackage{INTERSPEECH2020}
\usepackage{xcolor}
\usepackage{multirow}

\title{Waveform-based Voice Activity Detection Exploiting Fully Convolutional networks with Multi-Branched Encoders}
\name{Cheng Yu$^1$, Kuo-Hsuan Hung$^1$, I-Fan Lin$^1$, Szu-Wei Fu$^1$, Yu Tsao$^1$, Jeih-weih Hung$^2$}
\address{
  $^1$Research Center for Information Technology Innovation, Academia Sinica, Taiwan\\
  $^2$Department of Electrical Engineering, National Chi Nan University, Taiwan}
\email{r06943106@g.ntu.edu.tw, yu.tsao@citi.sinica.edu.tw}

\begin{document}

\maketitle
\begin{abstract}
  In this study, we propose an encoder-decoder structured system with fully convolutional networks to implement voice activity detection (VAD) directly on the time-domain waveform. The proposed system processes the input waveform to identify its segments to be either speech or non-speech. This novel waveform-based VAD algorithm, with a short-hand notation ``WVAD'', has two main particularities. First, as compared to most conventional VAD systems that use spectral features, raw-waveforms employed in WVAD contain more comprehensive information and thus are supposed to facilitate more accurate speech/non-speech predictions. Second, based on the multi-branched architecture, WVAD can be extended by using an ensemble of encoders, referred to as WEVAD, that incorporate multiple attribute information in utterances,  and thus can yield better VAD performance for specified acoustic conditions. We evaluated the presented WVAD and WEVAD for the VAD task in two datasets: First, the experiments conducted on AURORA2 reveal that WVAD outperforms many state-of-the-art VAD algorithms. Next, the TMHINT task confirms that through combining multiple attributes in utterances, WEVAD behaves even better than WVAD.

\end{abstract}

\section{Introduction}

Voice activity detection (VAD) aims to detect speech segments from audio streams and has been widely applied as a fundamental yet crucial front-end unit in various speech-related applications, such as automatic speech recognition (ASR) \cite{ramirez2007voice, ramirez2005statistical}, communication systems \cite{beritelli1998robust, lee1998voice}, speaker recognition systems \cite{alam2014supervised, mak2014study}, and speech enhancement (SE) \cite{boll1979suppression, lu2008geometric}. Conventional VAD methods are designed based on assumptions of speech and noise characteristics. One class of VAD identifies speech segments based on energy levels. Some notable examples are \cite{haigh1993robust, tanyer2000voice, woo2000robust, enqing2002voice, hsu2015robust}. Although the energy-level-based approaches can perform well in clean conditions, their detection precision is usually dropped when background noises are significant. Another class of VAD is established based on statistical models. For example, Sohn proposed an hidden Markov model (HMM) with a hang-over scheme \cite{sohn1999statistical} and Nemer applied higher-order cumulants of linear predictive coding (LPC) residuals \cite{nemer2001robust}. Later, Chang adopts probability distribution functions (pdf) for adaptive noise modeling \cite{chang2006voice}. Tan proposes a two-pass approach, which combines speech enhancement (SE) and VAD \cite{tan2020rvad}. Generally speaking, these conventional VAD approaches either directly estimate speech segments based on energy levels of speech signal, or isolate noise components from speech sequences with an adaptive/accurate noise tracking scheme. However, the underlying noise tracking schemes could not perform well under the time-varying noises because the assumptions of speech and noise properties do not necessarily hold. 

Owing to recent advances in machine learning (ML) and deep learning (DL), various ML/DL-based models have been used as a core unit in the VAD systems and achieved remarkable improvements over conventional approaches. As for the ML-based VAD approaches, the support vector machine (SVM) is usually served as the core module. Notable examples are \cite{enqing2002applying, jo2009statistical, shin2010voice, wu2011efficient}. In contrast, Some VAD algorithms are based on a Gaussian mixture model (GMM) framework. For example, in \cite{ying2011voice} Ying proposed a sequential Gaussian mixture model using expectation-maximization (EM) for initialization to achieve VAD. Compared with the ML-based methods that usually use hand-crafted features, the DL-based methods can exploit more extensive and versatile feature representations dwelled in the latent spaces (layers) of the employed deep neural network, such as recurrent neural networks (RNNS) \cite{hughes2013recurrent, tao2017bimodal} and long-short term memory (LSTM) \cite{eyben2013real, kim2018voice}. For example, in \cite{zhang2012deep} Zhang  proposed a fusion of multiple features using a deep belief network (DBF) that outperforms several past state-of-the-art works with a compact framework. Nevertheless, the respective VAD accuracy is not necessarily satisfactory when tested under some challenging noise situations.

\par
In light of the progress of VAD algorithms mentioned earlier, multiple features seem to be a better choice than a single feature to achieve superior speech/non-speech classification. In addition, adopting a deep-learning model in a VAD process has been one of the most favorable ways that can assure acceptable VAD accuracy and exhibit robustness against challenging noise environments. Therefore, in this study, we aim to design a novel VAD scenario that adopts multiple speech attributes and leverages a DL-based ensemble framework. Preliminary experiments have shown that this new waveform-based VAD method outperforms some state-of-the-art VAD algorithms for the Aurora2 dataset \cite{hirsch2000aurora}, and we have also shown that employing more speech attributes as the input to this new system benefits the VAD performance for the TMHINT database \cite{TMHINT}.

\graphicspath{ {./images/} }
\begin{figure}[t]
\centering
   \includegraphics[width=8cm]{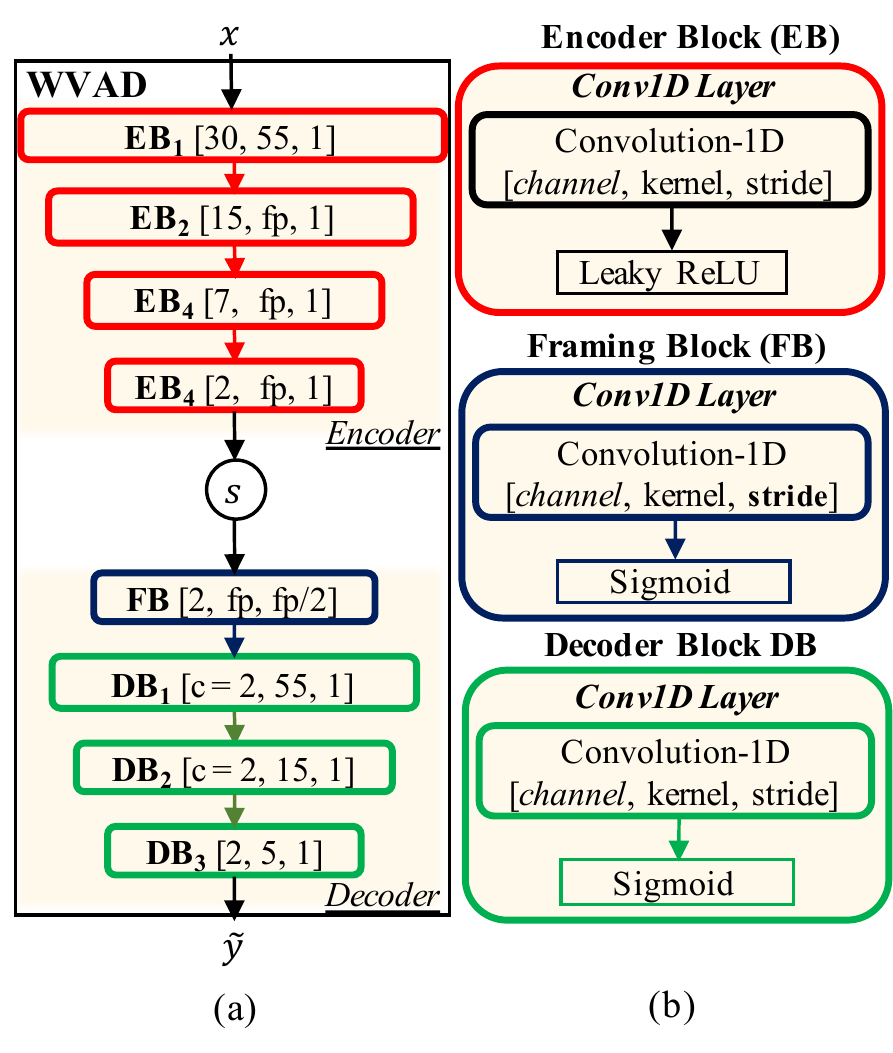} 
   \vspace{-0.4cm}
    \caption{The architecture of the porposed WVAD.}
    \label{fig:EDFCN}
    \vspace{-0.3cm}
\end{figure}

\section{Related Works}
This section briefly introduces some VAD algorithms that motivate us to develop the new VAD scenario.
\subsection{VAD using multiple features}
Some ML/DL-based VAD approaches shed light on the advantage of utilizing multiple acoustic features, like the work in \cite{wu2011efficient}. To further yield advantages from multiple features,  a deep ensemble framework with acoustic environment detection (AEC) is proposed in \cite{hwang2016ensemble}  that aims to explore benefits for VAD from multiple environmental features. Dey  proposed an ensemble framework based on SVM \cite{dey2018ensemble} that shows huge improvements compared to a single SVM for VAD. Zhang proposed an ensemble of classifiers (multi-resolution stack, MRS) \cite{zhang2015boosting} based on their proposed bDNN model. Noticeably, the MRS outperforms other VAD approaches by great margins in the area under curve (AUC) metric. However, these approaches generally use segmented spectral magnitude features, or power magnitude features from time-frequency analyses, where the phase information is not taken into consideration.
\subsection{VAD based on time-domain or phase-aware scheme}
Recently, phase features from time-frequency analyses have been shown to be unignorable for an effective VAD, according to what Wang  proposed in their phase-aware framework \cite{wang2017phase}. For even better use of full information from speech signal, Zazo  proposed a CLDNN framework \cite{zazo2016feature} that directly performs VAD on segmented waveform utterances. The convolutional blocks in CLDNN dissolve time-domain features as filter banks, and feeds the extracted features to a LSTM-DNN block for finalization of detection. However, the model complexity of CLDNN can sometimes be great since it requires the integration of three totally different neural network structures. 

\section{Proposed Method}

To integrate the advantages of the aforementioned approaches while maintaining a compact structure, we propose a waveform-based VAD framework using a encoder-decoder structure consisting of fully convolutional network, and we further improve this framework by integrating multiple speech attributes through an ensemble of encoders. 

\vspace{-0.15cm}
\graphicspath{ {./images/} }
\begin{figure}[t]
\centering
   \includegraphics[width=7.6cm]{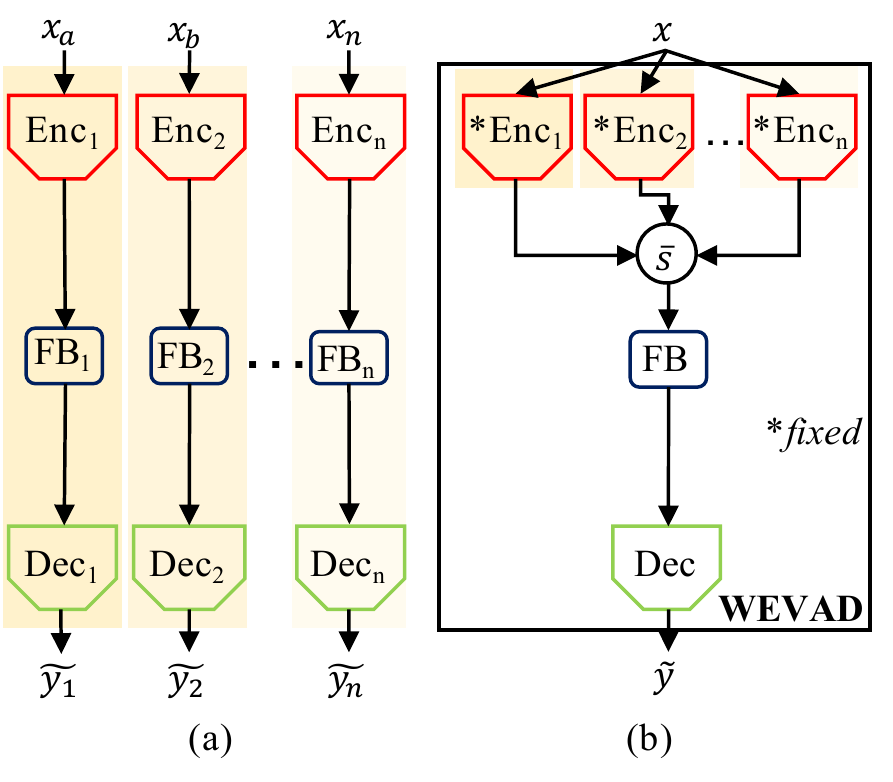} 
   \vspace{-0.4cm}
    \caption{The structure of the proposed WEVAD framework, using UAT-guided encoders.}
    \label{fig:EDEnsemble}
    \vspace{-0.3cm}
\end{figure}

\subsection{Encoder-Decoder FCN framework}
Partially inspired by  our previous works \cite{fu2018end}, this newly presented scenario primarily adopts an encoder-decoder (ED) scheme, which resorts to a fully convolutional network (FCN) to implement VAD directly on the time-domain speech waveforms. This novel waveform-based VAD algorithm, with a short-hand notation ``WVAD'', is depicted in Figure \ref{fig:EDFCN}(a), consists of an encoder of four convolutional blocks, a framing block, and a decoder of three convolutional blocks, which detailed arrangements are shown in Figure \ref{fig:EDFCN}(b).\\
Briefly speaking, WVAD converts the input waveform $x(t)$ to a series of frame-wise VAD output, $\tilde{y}=[y_s[\tau], y_{ns}[\tau]]$, where $y_s[\tau]$ and $y_{ns}[\tau]$ refers to the  scores for speech channel and non-speech channel, respectively, and $\tau$ is the frame index. For any specific frame $\tau$ in $x(t)$, if $y_s[\tau] \geq y_{ns}[\tau]$, it is predicted as a speech frame; otherwise it is predicted as non-speech. In the following sub-sections, we detail each portion of WVAD.

\vspace{-0.2cm}
\subsubsection{Encoder Framework}
Each of the four encoder blocks (EBs) shown in Figure \ref{fig:EDFCN}(a) consists of a one-dimensional convolutional layer followed by a leaky-ReLU activation function, as depicted in the upper part of Figure \ref{fig:EDFCN}(b). These four EBs are stacked in an inverted-triangular fashion as in Figure \ref{fig:EDFCN}(a) (by decreasing the number of channels) to downscale the feature map into two channels, denoted by $s$.

\vspace{-0.2cm}
\subsubsection{Framing Block (FB) framework}

Since the VAD targets are labeled on the frames and stride half of frame points each label (that is, the speech/non-speech labeling is frame-based and the neighboring frames have a 50\% overlap), we thus implement a framing block (FB) which adopts a one-dimensional layer with the same stride as the VAD labeling. As shown in the middle part of  Figure\ref{fig:EDFCN}(b), the FB consists of a  convolution layer, which receives the feature map $s$ from the preceding encoder framework, and a sigmoid activation layer. Briefly speaking, here the FB serves as a primitive speech-nonspeech classifier, in which the two channels in the convolutional layer attempt to reflect the speech and non-speech frames, respectively. 


\vspace{-0.2cm}
\subsubsection{Decoder Framework}
Here the decoder framework is to further polish up the output of the FB in order to achieve a superior VAD performance. This decoder framework is created through stacking decoder blocks (DBs), each consisting of a two-channel convolutional layer and a sigmoid activation layer, as shown in the lower part of Figure\ref{fig:EDFCN}(b). Similar to the encoder framework, here these DBs are stacked in an inverted triangular fashion. In these DBs (denoted by DB$_{1}$, DB$_{2}$, and DB$_{3}$) each convolutional layer resolves time-frequency representations in the channels of feature maps, while the shrinking kernel sizes $[55, 15, 5]$
reflect the decrease of modulation frequency resolution from 1.83 Hz to 20 Hz (with the segment rate being 100 Hz). The operations involved in the convolutional layers can be expressed by

\vspace{-0.2cm}
\begin{gather}
z_{1, m}(\tau) \ = \ \sigma (\mathbf{F}_{1, m}^{T} \circledast z_{1,m-1}(\tau)+\mathbf{b}_{1, m}),  \\
z_{2, m}(\tau) \ = \ \sigma (\mathbf{F}_{1, m}^{T} \circledast z_{2,m-1}(\tau)+\mathbf{b}_{2, m}),  
\end{gather}
where  $\mathbf{F}_{j, m}$ $\in \mathbb{R}^{ L \times I}$ and $\mathbf{b}_{1, m}$ denote the filter (kernel) and bias for channel $j$ in a convolutional layer of DB$_{m}$ respectively, $\tau$ denotes segment time index in each channel, $z_{j,m}(\tau)$ denotes the feature map of channel $j$ from DB$_{m}$, $m =1, 2, 3$, and   $z_{j,0}(\tau)$ is the feature map of channel $j$ from the FB framework.
Thus, given the input speech waveform $x(t)$, we have the following outputs in turn:
\vspace{-0.1cm}
\begin{gather}
\mathbf{s}(t) \ = \ [s_{1}(t); s_{2}(t)] \ = \ \textbf{Encoder}(x(t)), \\
\mathbf{z}_{0}(\tau) \ =  \ [z_{1,0}(\tau); z_{2,0}(\tau)] =\textbf{FB}(\mathbf{s}(t)),\ \\
\mathbf{\tilde{y}}(\tau) \ =  \ [z_{1, 3}(\tau); z_{2,3}(\tau)] =\textbf{DB}_3(\textbf{DB}_2(\textbf{DB}_1(\mathbf{z}_{0}(\tau)),
\end{gather}
where $z_{1, 3}(\tau)$ and $z_{2, 3}(\tau)$ correspond to the speech channel and non-speech channel, respectively, in the predicted VAD label vector $\mathbf{\tilde{y}}(\tau)$. 

\subsection{Encoders with Utterance-Level Attribute Tree (UAT)}

Here, we further propose to refine WVAD by exploiting multiple encoders by means of an utterance-level attribute tree (UAT). According to our recent study \cite{yu2020speech}, a UAT divides the utterances in the training set into several subsets according to different attributes. For example, let a UAT adopt two levels of attribute, namely gender and signal-to-noise ratio (SNR). The first level consists two nodes corresponding to two genders: male and female, which subsequently makes four nodes in the second level, namely male and high-SNR, female and high-SNR, male and low-SNR, and female and low-SNR. As a result, an extension of WVAD that adopts an ensemble of attributes,  abbreviated as WEVAD, can be created and operated with the following procedures:\\
\textbf{Step 1}: We use a UAT to split the training data into several subsets, each subset belonging to a particular attribute or the intersection of several attributes. Then for each subset, an WVAD model is trained, as shown in Fig. \ref{fig:EDEnsemble}(a).\\
\textbf{Step 2}: The encoder part of all UAT-guided attribute-wise WVAD models is collected and arranged in parallel to be an ensemble of encoders, which outputs are concatenated and then fed to a frame block (FB) network and a decoder network in turn, as depicted in Fig \ref{fig:EDEnsemble}(b). This new framework is trained using the whole training set, while only the parameters in the FB and decoder frameworks are learned and the encoder ensemble is frozen. \\
\textbf{Step 3}: In the test phase, an arbitrary test utterance $x$  is first fed to the encoder ensemble to produce multiple feature maps, which are concatenated as follows:
\begin{equation}
\tilde{s} \ = \ [\textbf{Encoder}_{1}(x), \textbf{Encoder}_{2}(x), ..., \textbf{Encoder}_{n}(x)],     
\end{equation}
where $n$ is the number of encoders in the ensemble. Finally, the feature map ensemble $\tilde{s}$ is passed through the subsequent FB and decoder frameworks to obtain the predicted VAD label sequence $\mathbf{\tilde{y}}(\tau)$, as described in Eq. (5).


\begin{table*}[t]\scriptsize
\centering
\caption{Accuracy (\%) of WVAD and several state-of-the-art approaches at 4 noise types (Restaurant, Street, Airport, Train), 1 channel mismatch noise type (Subway) at four SNR levels in AURORA2.}
\vspace{-0.5cm}
\begin{center} 
\begin{tabular}{|c|c|c|c|c|c|c|c|c||c|}
\hline
\textbf{Noise type} & \textbf{SNR} & \textbf{G.729B\cite{itu1996silence}} & \textbf{Sohn\cite{sohn1999statistical}} & \textbf{Ramirez05\cite{ramirez2005statistical}} & \textbf{Yu\cite{yu2010discriminative}} & \textbf{MK-SVM\cite{wu2011efficient}} & \textbf{Zhang13\cite{zhang2012deep}} & \textbf{AEC\cite{hwang2016ensemble}} & \textbf{WVAD} \\ \hline
\multirow{4}{*}{\textbf{Restaurant}} & -5dB & 57.76 & 64.38 & 64.38 & 64.38 & 70.44 & 70.10 & 82.91 & \textbf{86.99}\\
& 0dB & 65.31 & 64.38 & 64.56 & 64.51 & 75.71 & 75.68 & 89.57 & \textbf{91.04}\\ 
& 5dB & 69.67 & 66.03 & 69.59 & 68.10 & 83.25 & 83.59 & 93.01 & \textbf{94.31}\\
& 10dB & 72.46 & 70.02 & 75.65 & 73.38 & 86.30 & 86.08 & 95.09 & \textbf{95.72}\\
\hline
\multirow{4}{*}{\textbf{Street}} & -5dB & 57.45 & 54.58 & 55.25 & 54.58 & 63.38 & 67.41 & 81.50 & \textbf{86.89}\\
& 0dB & 65.71 & 57.43 & 58.28 & 57.59 & 73.35 & 73.76 & 89.57 & \textbf{91.17}\\ 
& 5dB & 72.63 & 64.84 & 67.69 & 65.68 & 77.60 & 78.70 & 93.07 & \textbf{94.27}\\
& 10dB & 74.45 & 70.07 & 69.52 & 71.05 & 79.10 & 80.86 & 94.41 & \textbf{95.81}\\
\hline
\multirow{4}{*}{\textbf{Airport}} & -5dB & 57.00 & 56.94 & 57.18 & 57.53 & 65.86 & 66.35 & 86.35 & \textbf{87.96}\\
& 0dB & 65.54 & 61.32 & 62.22 & 62.29 & 75.59 & 76.66 & 91.33 & \textbf{91.95}\\ 
& 5dB & 69.64 & 68.25 & 71.46 & 70.21 & 82.30 & 81.92 & 94.22 & \textbf{94.54}\\
& 10dB & 72.02 & 77.31 & 80.05 & 80.04 & 85.38 & 86.41 & 94.96 & \textbf{95.89}\\
\hline
\multirow{4}{*}{\textbf{Train}} & -5dB & 57.56 & 58.32 & 58.41 & 58.20 & 68.78 & 68.99 & \textbf{89.48} & 88.27\\
& 0dB & 67.91 & 59.48 & 61.17 & 59.95 & 76.31 & 76.95 & \textbf{93.42} & 92.37\\ 
& 5dB & 75.26 & 68.84 & 72.89 & 70.88 & 83.99 & 83.49 & 94.99 & \textbf{95.21}\\
& 10dB & 77.05 & 75.81 & 79.35 & 78.42 & 85.34 & 85.68 & 95.64 & \textbf{96.14}\\
\hline
\multirow{4}{*}{\textbf{Subway}} & -5dB & 49.25 & 68.23 & 68.15 & 68.25 & 79.50 & 79.10 & 75.94 & \textbf{89.78}\\
& 0dB & 55.20 & 68.15 & 68.16 & 68.18 & 83.28 & 83.29 & 84.46 & \textbf{93.76}\\ 
& 5dB & 62.08 & 68.64 & 73.16 & 69.68 & 86.11 & 85.77 & 90.80 & \textbf{95.30}\\
& 10dB & 70.51 & 70.03 & 77.93 & 72.93 & 87.46 & 86.25 & 93.99 & \textbf{95.92}\\

\hline
\end{tabular}
\end{center}
\label{tab:ED-FCN-Acc}
\vspace{-0.69cm}
\end{table*}

%

\begin{table}[ht]\scriptsize
\centering
\caption{Accuracy (\%) averaged over 7 noise types at six SNR levels in AURORA2.}
\vspace{-0.5cm}
\begin{center} 
\begin{tabular}{|p{1.84cm}|p{0.36cm}p{0.36cm}p{0.36cm}p{0.36cm}p{0.36cm}p{0.496cm}|p{0.5cm}|}
 \hline
 \multirow{2}{*}{\textbf{Methods}} & \multicolumn{7}{c|}{AUC(\%)} \\
 \cline{2-8}
 & 20dB & 15dB & 10dB & 5dB & 0dB & \textbf{-}5dB & \textbf{AVG.} \\
 \hline 
 \textbf{MRCG-SVM\cite{zhang2015boosting}} & 95.13 & 94.16 & 92.55 &  89.93 & 83.74 & 75.50 & 88.50\\
 \textbf{Zhang13\cite{zhang2012deep}} & 95.93 & 95.32 & 93.86 & 91.25 & 85.12 & 77.23 & 89.79\\
 \textbf{bDNN\cite{zhang2015boosting}} & 96.09 & 95.64 & 94.97 & 93.86 & 90.92 & 85.94 & 92.90\\
 \textbf{MRS\cite{zhang2015boosting}} & 96.56 & 96.15 & 95.58 & 94.56 & 91.91 & 87.20 & 93.66\\
 \textbf{WVAD} & \textbf{99.45} & \textbf{99.40} & \textbf{99.27} & \textbf{98.80} & \textbf{97.04} & \textbf{92.29} & \textbf{97.71}\\
 \hline
\end{tabular}
\end{center}
\label{tab:ED-FCN-AUC}
\vspace{-0.5cm}
\end{table}

\section{Experimental Setup and Results}

Here, we first examine the presented WVAD by comparing it with several state-of-the-art VAD algorithms in terms of Accuracy and AUC. This comparison is implemented on the AURORA2 corpus \cite{hirsch2000aurora} with the VAD labeling in \cite{tan2010low}, which contains 8,440 and 60,060 utterances at a sampling rate of 8k Hz for multi-condition training and testing, respectively.  Next, we use the Taiwan Mandarin version of the hearing in noise test (TMHINT) \cite{TMHINT} corpus to evaluate WVAD, by varying the settings of the attribute in the used UAT. MHINT contains 36,000 training utterances that are artificially corrupted with 100 noise types at 31 different SNRs (from -10dB to 20dB).
For TMHINT, the target frame-based VAD labels are created by applying the rVAD \cite{tan2020rvad} algorithm to clean noise-free utterances. 
Note that the VAD label of each frame is organized as a vector of two channels. The non-speech frame is labeled as $[1,0]$, while the speech frame is labeled as $[0,1]$.

In addition, with the TMHINT dataset, two versions of WEVAD, WEVAD$_{(2)}$ and WEVAD$_{(6)}$, are evaluated and compared, in which  WEVAD$_{(2)}$ corresponds to 2 gender nodes of encoders (male and female) and WEVAD$_{(6)}$ is associated with 6 nodes of encoders (male, male high-SNR, male low-SNR, female, female high-SNR, and female low-SNR).

Notably, we report the VAD results in two metrics: ACC(\%) and AUC(\%). ACC indicates the VAD accuracy, calculated as the ratio of the number of correctly predicted frames to the number of total frames, and AUC \cite{zhang2015boosting} is the abbreviation of ``area under the curve'',  which involves the receiver operating characteristic (ROC) curve with the x-axis for false-positive rate (FPR) and the y-axis for the true-positive rate (TPR).

\subsection{Effects and performances of WVAD on AURORA2}

At the outset, Tables \ref{tab:ED-FCN-Acc} and \ref{tab:ED-FCN-AUC} list the ACC and AUC scores, respectively, for the presented WVAD and several state-of-the-art VAD algorithms for the AURORA2 corpus. It is clearly shown in Table\ref{tab:ED-FCN-Acc} that WVAD provides the optimal ACC scores for almost all noise types and levels (except for the train noise at -5 dB and 0 dB SNR) among the eight methods. In addition, from Table \ref{tab:ED-FCN-AUC} WVAD achieves significantly higher AUC scores than the other four algorithms at all SNR levels. These results demonstrate that WVAD behave quite excellent as well as robust for the VAD task under a wide range of noise environments.   

\begin{figure}[ht]
\centering
   \includegraphics[width=8cm, height=2.5cm]{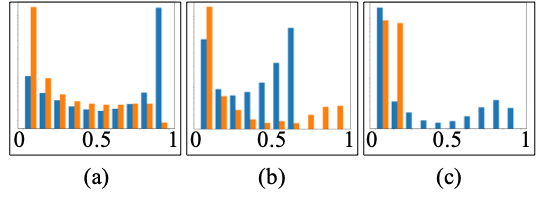}
   \vspace{-0.4cm}
    \caption{The output histograms of the two channels for (a)DB$_{1}$, (b) DB$_{2}$, and (c) DB$_{3}$ in log scale (y-axis), with the -5dB-SNR babble test set in AURORA2.}
    \label{fig:DB_out}
    \vspace{-0.4cm}
\end{figure}

\noindent Next, Figure \ref{fig:DB_out}(a)(b)(c) shows the output histograms of three decoder blocks (DB$_1$, DB$_2$ and DB$_3$) of WVAD shown in Figure \ref{fig:EDFCN}(a) for the test set with babble noise at -5 dB SNR, where the blue bars and orange bars refer to the non-speech channel and speech channel, respectively. From Figures\ref{fig:DB_out}(a) to (c), we find that the histogram for the speech channel (with orange color) gradually concentrates into two groups with low and high values, indicating its increasing discriminative ability. 

\begin{figure}[ht]
\vspace{-0.1cm}
\centering
   \includegraphics[width=8cm,height=2.8cm]{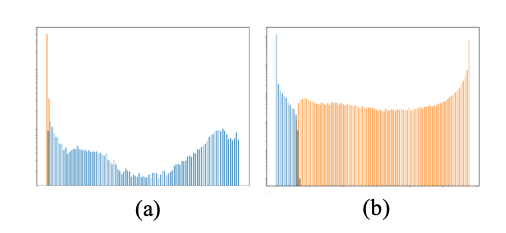}
   \vspace{-0.6cm}
    \caption{the histograms of two channels with respect to (a) true-negative (TN) and (b) true-positive (TP) cases of the final output $y$ in log scale (y-axis), with the -5dB-SNR babble test set in AURORA2.}
    \label{fig:tptn}
\vspace{-0.25cm}
\end{figure}

\noindent Finally, Figures \ref{fig:tptn}(a) and (b) show the histograms with respect to true-negative (TN) and true-positive (TP) cases of the final output $\tilde{y}$ for the same test set as Figure \ref{fig:DB_out}(c). The two figures evidently show that the proposed WVAD can help establish more distinctive numerical distribution from the speech channel (with orange color) than the non-speech channel (with blue color).

\graphicspath{ {./images/} }

\graphicspath{ {./images/} }

\subsection{Results for the TMHINT corpus}

\begin{figure}[ht]
\centering
   \includegraphics[width=6cm]{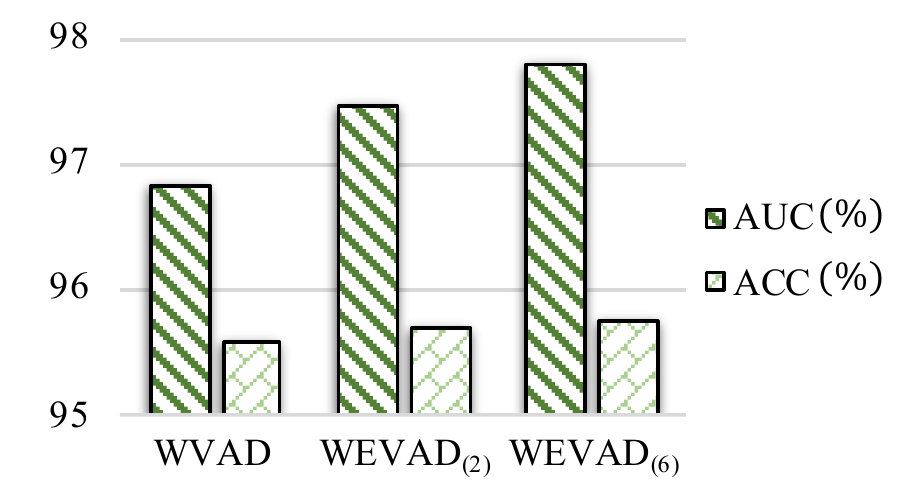}
   \vspace{-0.4cm}
    \caption{The average accuracy and AUC in TMHINT over 4 noise tpyes at 6 SNR levels using WVAD, WEVAD$_{(2)}$ and WEVAD$_{(6)}$.}
    \label{fig:AA}
\end{figure}

In this section, we examine the performances of WEVAD, WEVAD$_{(2)}$ and WEVAD$_{(6)}$ for the TMHINT task. Considering the fact that the sampling rate of the utterances in TMHINT is 16 kHz,  different from that of AURORA2, we slightly adjust the settings of convolutional kernels and strides to fit the corpus. We prepared 5,760 test utterances with four noise types (babble, car, engine and street) that were unseen
by the training stage to compare the three frameworks. The respective ACC and AUC results are listed in Table \ref{tab:3c-ACCAUC} and further summarized in Figure 
\ref{fig:AA}. From Figure \ref{fig:AA}, we see that all the three frameworks can achieve over 94\% in averaged ACC and AUC scores, while adopting multiple encoders as in WEVAD$_{(2)}$ and WEVAD$_{(6)}$ benefit AUC scores significantly and provide a moderate ACC improvement relative to WVAD. Further examining Table \ref{tab:3c-ACCAUC} we find that WEVAD$_{(6)}$ obtains the optimal AUC scores at all SNR situations, and performs better in ACC metric for most SNR cases. These results clearly confirm the effectiveness of the presented multiple-encoder frameworks to implement VAD in various noisy situations.



\graphicspath{ {./images/} }

\begin{table}[ht]\scriptsize
\centering
\caption{\scshape{A}UC(\%) \& ACC(\%) average over 4 noise types at six SNR levles in TMHINT.}
\vspace{-0.6cm}
\begin{center} 
\begin{tabular}{|p{1.4cm}|p{0.36cm}p{0.36cm}p{0.36cm}p{0.36cm}p{0.496cm}p{0.65cm}|p{0.5cm}|}
 \hline
 \multirow{2}{*}{\textbf{Methods}} & \multicolumn{7}{c|}{AUC(\%)} \\
 \cline{2-8}
 & 15dB & 10dB & 5dB & 0dB & \textbf{-}5dB & \textbf{-}10dB & \textbf{AVG.} \\
 \hline 
 \textbf{WVAD} & 98.84 & 98.82 & 98.71 & 98.16 & 95.87 & 90.62 & 96.84\\
 \textbf{WEVAD$_{(2)}$} & 99.21 & 99.12 & 98.89 & 98.20 & 96.35 & 92.94 & 97.54\\
 \textbf{WEVAD$_{(6)}$} & \textbf{99.25} & \textbf{99.26} & \textbf{99.15} & \textbf{98.59} & \textbf{96.71} & \textbf{93.73} & \textbf{97.78}\\
 \hline
\end{tabular}
\begin{tabular}{|p{1.4cm}|p{0.36cm}p{0.36cm}p{0.36cm}p{0.36cm}p{0.496cm}p{0.65cm}|p{0.5cm}|}
 \hline
 \multirow{2}{*}{\textbf{Methods}} & \multicolumn{7}{c|}{ACC(\%)} \\
 \cline{2-8}
 & 15dB & 10dB & 5dB & 0dB & \textbf{-}5dB & \textbf{-}10dB & \textbf{AVG.} \\
 \hline 
  \textbf{WVAD} & 97.79 & 97.54 & 97.08 & 95.86 & \textbf{93.86} & 91.33 & 95.58\\
  \textbf{WEVAD$_{(2)}$} & 97.82 & 97.59 & 97.17 & 95.88 & 93.64 & \textbf{91.97} & 95.68\\
  \textbf{WEVAD$_{(6)}$} & \textbf{97.87} & \textbf{97.74} & \textbf{97.50} & \textbf{96.20} & 93.63 & 91.54 & \textbf{95.75}\\
 \hline
\end{tabular}
\end{center}
\label{tab:3c-ACCAUC}
\vspace{-0.9cm}
\end{table}

\section{Conclusion}

In this study, we present novel waveform-based VAD scenarios that leverage an encoder-decoder structure consisting of fully convolutional networks. Furthermore, by virtue of the utterance-level attribute tree (UAT) algorithm, we can create an ensemble of attribute-wise encoders in the leading part of the whole VAD system, which is believed to capture more beneficial VAD information than a single encoder. Experiments conducted on the AURORA2 dataset reveals that the presented WVAD behaves significantly better than many state-of-the-art VAD algorithms, and through the TMHINT evaluation results we further validate that the encoder-ensemble version of WVAD, viz. WEVAD,  can further promote the VAD performance in terms of ACC and AUC metric scores. We conclude that the newly presented methods can provide accurate VAD results for utterances distorted by noise of various types and SNR levels.


\bibliographystyle{IEEEtran}

\bibliographystyle{IEEEbib}


\begin{thebibliography}{10}
\providecommand{\url}[1]{#1}
\csname url@samestyle\endcsname
\providecommand{\newblock}{\relax}
\providecommand{\bibinfo}[2]{#2}
\providecommand{\BIBentrySTDinterwordspacing}{\spaceskip=0pt\relax}
\providecommand{\BIBentryALTinterwordstretchfactor}{4}
\providecommand{\BIBentryALTinterwordspacing}{\spaceskip=\fontdimen2\font plus
\BIBentryALTinterwordstretchfactor\fontdimen3\font minus
  \fontdimen4\font\relax}
\providecommand{\BIBforeignlanguage}[2]{{%
\expandafter\ifx\csname l@#1\endcsname\relax
\typeout{** WARNING: IEEEtran.bst: No hyphenation pattern has been}%
\typeout{** loaded for the language `#1'. Using the pattern for}%
\typeout{** the default language instead.}%
\else
\language=\csname l@#1\endcsname
\fi
#2}}
\providecommand{\BIBdecl}{\relax}
\BIBdecl

\bibitem{ramirez2007voice}
J.~Ramirez, J.~M. G{\'o}rriz, and J.~C. Segura, ``Voice activity detection.
  fundamentals and speech recognition system robustness,'' \emph{Robust speech
  recognition and understanding}, vol.~6, no.~9, pp. 1--22, 2007.

\bibitem{ramirez2005statistical}
J.~Ram{\'\i}rez, J.~C. Segura, C.~Ben{\'\i}tez, L.~Garc{\'\i}a, and A.~Rubio,
  ``Statistical voice activity detection using a multiple observation
  likelihood ratio test,'' \emph{IEEE Signal Processing Letters}, vol.~12,
  no.~10, pp. 689--692, 2005.

\bibitem{beritelli1998robust}
F.~Beritelli, S.~Casale, and A.~Cavallaero, ``A robust voice activity detector
  for wireless communications using soft computing,'' \emph{IEEE Journal on
  Selected Areas in Communications}, vol.~16, no.~9, pp. 1818--1829, 1998.

\bibitem{lee1998voice}
I.~D. Lee, H.~P. Stern, and S.~Mahmoud, ``A voice activity detection algorithm
  for communication systems with dynamically varying background acoustic
  noise,'' in \emph{Proc. VTC}, vol.~2.\hskip 1em plus 0.5em minus 0.4em\relax
  IEEE, 1998, pp. 1214--1218.

\bibitem{alam2014supervised}
J.~Alam, P.~Kenny, P.~Ouellet, T.~Stafylakis, and P.~Dumouchel,
  ``Supervised/unsupervised voice activity detectors for text-dependent speaker
  recognition on the rsr2015 corpus,'' in \emph{Proc. Odyssey. Workshop}, 2014,
  pp. 123--130.

\bibitem{mak2014study}
M.-W. Mak and H.-B. Yu, ``A study of voice activity detection techniques for
  nist speaker recognition evaluations,'' \emph{Computer Speech \& Language},
  vol.~28, no.~1, pp. 295--313, 2014.

\bibitem{boll1979suppression}
S.~Boll, ``Suppression of acoustic noise in speech using spectral
  subtraction,'' \emph{IEEE Transactions on acoustics, speech, and signal
  processing}, vol.~27, no.~2, pp. 113--120, 1979.

\bibitem{lu2008geometric}
Y.~Lu and P.~C. Loizou, ``A geometric approach to spectral subtraction,''
  \emph{Speech communication}, vol.~50, no.~6, pp. 453--466, 2008.

\bibitem{haigh1993robust}
J.~Haigh and J.~Mason, ``Robust voice activity detection using cepstral
  features,'' in \emph{Proc. TENCON}, vol.~3.\hskip 1em plus 0.5em minus
  0.4em\relax IEEE, 1993, pp. 321--324.

\bibitem{tanyer2000voice}
S.~G. Tanyer and H.~Ozer, ``Voice activity detection in nonstationary noise,''
  \emph{IEEE Transactions on speech and audio processing}, vol.~8, no.~4, pp.
  478--482, 2000.

\bibitem{woo2000robust}
K.-H. Woo, T.-Y. Yang, K.-J. Park, and C.~Lee, ``Robust voice activity
  detection algorithm for estimating noise spectrum,'' \emph{Electronics
  Letters}, vol.~36, no.~2, pp. 180--181, 2000.

\bibitem{enqing2002voice}
D.~Enqing, L.~Guizhong, Z.~Yatong, and C.~Yu, ``Voice activity detection based
  on short-time energy and noise spectrum adaptation,'' in \emph{Proc. ICOSP},
  vol.~1.\hskip 1em plus 0.5em minus 0.4em\relax IEEE, 2002, pp. 464--467.

\bibitem{hsu2015robust}
C.-C. Hsu, K.-M. Cheong, T.-S. Chi, and Y.~Tsao, ``Robust voice activity
  detection algorithm based on feature of frequency modulation of harmonics and
  its dsp implementation,'' \emph{IEICE Transactions on Information and
  Systems}, vol.~98, no.~10, pp. 1808--1817, 2015.

\bibitem{sohn1999statistical}
J.~Sohn, N.~S. Kim, and W.~Sung, ``A statistical model-based voice activity
  detection,'' \emph{IEEE signal processing letters}, vol.~6, no.~1, pp. 1--3,
  1999.

\bibitem{nemer2001robust}
E.~Nemer, R.~Goubran, and S.~Mahmoud, ``Robust voice activity detection using
  higher-order statistics in the lpc residual domain,'' \emph{IEEE Transactions
  on Speech and Audio Processing}, vol.~9, no.~3, pp. 217--231, 2001.

\bibitem{chang2006voice}
J.-H. Chang, N.~S. Kim, and S.~K. Mitra, ``Voice activity detection based on
  multiple statistical models,'' \emph{IEEE Transactions on Signal Processing},
  vol.~54, no.~6, pp. 1965--1976, 2006.

\bibitem{tan2020rvad}
Z.-H. Tan, N.~Dehak \emph{et~al.}, ``rvad: An unsupervised segment-based robust
  voice activity detection method,'' \emph{Computer Speech \& Language},
  vol.~59, pp. 1--21, 2020.

\bibitem{enqing2002applying}
D.~Enqing, L.~Guizhong, Z.~Yatong, and Z.~Xiaodi, ``Applying support vector
  machines to voice activity detection,'' in \emph{Proc. ICSP-6}, vol.~2.\hskip
  1em plus 0.5em minus 0.4em\relax IEEE, 2002, pp. 1124--1127.

\bibitem{jo2009statistical}
Q.-H. Jo, J.-H. Chang, J.~Shin, and N.~Kim, ``Statistical model-based voice
  activity detection using support vector machine,'' \emph{IET Signal
  Processing}, vol.~3, no.~3, pp. 205--210, 2009.

\bibitem{shin2010voice}
J.~W. Shin, J.-H. Chang, and N.~S. Kim, ``Voice activity detection based on
  statistical models and machine learning approaches,'' \emph{Computer Speech
  \& Language}, vol.~24, no.~3, pp. 515--530, 2010.

\bibitem{wu2011efficient}
J.~Wu and X.-L. Zhang, ``Efficient multiple kernel support vector machine based
  voice activity detection,'' \emph{IEEE Signal Processing Letters}, vol.~18,
  no.~8, pp. 466--469, 2011.

\bibitem{ying2011voice}
D.~Ying, Y.~Yan, J.~Dang, and F.~K. Soong, ``Voice activity detection based on
  an unsupervised learning framework,'' \emph{IEEE Transactions on Audio,
  Speech, and Language Processing}, vol.~19, no.~8, pp. 2624--2633, 2011.

\bibitem{hughes2013recurrent}
T.~Hughes and K.~Mierle, ``Recurrent neural networks for voice activity
  detection,'' in \emph{Proc. ICASSP}.\hskip 1em plus 0.5em minus 0.4em\relax
  IEEE, 2013, pp. 7378--7382.

\bibitem{tao2017bimodal}
F.~Tao and C.~Busso, ``Bimodal recurrent neural network for audiovisual voice
  activity detection.'' in \emph{Proc. Interspeech}, 2017, pp. 1938--1942.

\bibitem{eyben2013real}
F.~Eyben, F.~Weninger, S.~Squartini, and B.~Schuller, ``Real-life voice
  activity detection with lstm recurrent neural networks and an application to
  hollywood movies,'' in \emph{Proc. ICASSP}.\hskip 1em plus 0.5em minus
  0.4em\relax IEEE, 2013, pp. 483--487.

\bibitem{kim2018voice}
J.~Kim and M.~Hahn, ``Voice activity detection using an adaptive context
  attention model,'' \emph{IEEE Signal Processing Letters}, vol.~25, no.~8, pp.
  1181--1185, 2018.

\bibitem{zhang2012deep}
X.-L. Zhang and J.~Wu, ``Deep belief networks based voice activity detection,''
  \emph{IEEE Transactions on Audio, Speech, and Language Processing}, vol.~21,
  no.~4, pp. 697--710, 2012.

\bibitem{hirsch2000aurora}
H.-G. Hirsch and D.~Pearce, ``The aurora experimental framework for the
  performance evaluation of speech recognition systems under noisy
  conditions,'' in \emph{Proc. ISCA ITRW ASR2000}, 2000.

\bibitem{TMHINT}
M.~Huang, ``Development of taiwan mandarin hearing in noise test,''
  \emph{Department of speech language pathology and audiology, National Taipei
  University of Nursing and Health Science}, 2005.

\bibitem{hwang2016ensemble}
I.~Hwang, H.-M. Park, and J.-H. Chang, ``Ensemble of deep neural networks using
  acoustic environment classification for statistical model-based voice
  activity detection,'' \emph{Computer Speech \& Language}, vol.~38, pp. 1--12,
  2016.

\bibitem{dey2018ensemble}
J.~Dey, M.~S.~B. Hossain, and M.~A. Haque, ``An ensemble svm-based approach for
  voice activity detection,'' in \emph{Proc. ICECE}.\hskip 1em plus 0.5em minus
  0.4em\relax IEEE, 2018, pp. 297--300.

\bibitem{zhang2015boosting}
X.-L. Zhang and D.~Wang, ``Boosting contextual information for deep neural
  network based voice activity detection,'' \emph{IEEE/ACM Transactions on
  Audio, Speech, and Language Processing}, vol.~24, no.~2, pp. 252--264, 2015.

\bibitem{wang2017phase}
L.~Wang, K.~Phapatanaburi, Z.~Go, S.~Nakagawa, M.~Iwahashi, and J.~Dang,
  ``Phase aware deep neural network for noise robust voice activity
  detection,'' in \emph{Proc. ICME}.\hskip 1em plus 0.5em minus 0.4em\relax
  IEEE, 2017, pp. 1087--1092.

\bibitem{zazo2016feature}
R.~Zazo, T.~N. Sainath, G.~Simko, and C.~Parada, ``Feature learning with
  raw-waveform cldnns for voice activity detection,'' \emph{Proc. Interspeech},
  pp. 3668--3672, 2016.

\bibitem{fu2018end}
S.-W. Fu, T.-W. Wang, Y.~Tsao, X.~Lu, and H.~Kawai, ``End-to-end waveform
  utterance enhancement for direct evaluation metrics optimization by fully
  convolutional neural networks,'' \emph{IEEE/ACM Transactions on Audio,
  Speech, and Language Processing}, vol.~26, no.~9, pp. 1570--1584, 2018.

\bibitem{yu2020speech}
C.~Yu, R.~E. Zezario, J.~Sherman, Y.-Y. Hsieh, X.~Lu, H.-M. Wang, and Y.~Tsao,
  ``Speech enhancement based on denoising autoencoder with multi-branched
  encoders,'' \emph{arXiv preprint arXiv:2001.01538}, 2020.

\bibitem{itu1996silence}
A.~Itu, ``silence compression scheme for g. 729 optimized for terminals
  conforming to recommendation v. 70,'' \emph{ITU-T Recommendation G}, vol.
  729, 1996.

\bibitem{yu2010discriminative}
T.~Yu and J.~H. Hansen, ``Discriminative training for multiple observation
  likelihood ratio based voice activity detection,'' \emph{IEEE Signal
  Processing Letters}, vol.~17, no.~11, pp. 897--900, 2010.

\bibitem{tan2010low}
Z.-H. Tan and B.~Lindberg, ``Low-complexity variable frame rate analysis for
  speech recognition and voice activity detection,'' \emph{IEEE Journal of
  Selected Topics in Signal Processing}, vol.~4, no.~5, pp. 798--807, 2010.

\end{thebibliography}


\end{document}